\documentclass[runningheads]{llncs}
\usepackage{hyperref}

\usepackage{graphicx}
\usepackage{listings}
\usepackage{xcolor}

\lstdefinestyle{my_style}{
  basicstyle=\small\ttfamily,
  keywordstyle=\color{blue},
  frame=single,
  breaklines=true,
  captionpos=b,
  language=C,
  morekeywords={size_t,ucs_status_t,uint64_t,ucp_worker_h,ucp_ifunc_h,ucp_ep_h,
                ucp_rkey_h,ucp_ifunc_msg_t},
}

\newcommand{\cc}{\textit{Two-Chains}}
\newcommand{\ifunc}{\texttt{ifunc}}
\newcommand{\ifuncs}{\texttt{ifuncs}}

\begin{document}
\title{UCX Programming Interface for Remote Function Injection and
       Invocation}
%
%
\author{Luis~E.~Peña\inst{1}\thanks{Contributed equally} \and
        Wenbin Lu\inst{2}\protect\footnotemark[1] \and
        Pavel Shamis\inst{1} \and
        Steve Poole\inst{3}}
\authorrunning{L. E. Peña et al.}

\institute{Arm Research, Austin TX 78735, USA\\
\email{\{Luis.EPena,Pavel.Shamis\}@arm.com} \and
Stony Brook University, Stony Brook NY 11794, USA\\
\email{Wenbin.Lu@stonybrook.edu} \and
Los Alamos National Laboratory, Los Alamos NM 87545, USA\\
\email{swpoole@lanl.gov}}

\maketitle              

\begin{abstract}
Network library APIs have historically been developed with the emphasis on data
movement, placement, and communication semantics. Many communication semantics
are available across a large variety of network libraries, such as send-receive,
data streaming, put/get/atomic, RPC, active messages, collective communication,
etc. In this work we introduce new compute and data movement APIs that overcome
the constraints of the single-program, multiple-data (SPMD) programming model by
allowing users to send binary executable code between processing elements. Our
proof-of-concept implementation of the API is based on the UCX communication
framework and leverages the RDMA network for fast compute migration. We envision
the API being used to dispatch user functions from a host CPU to a SmartNIC
(DPU), computational storage drive (CSD), or remote servers. In addition, the
API can be used by large-scale irregular applications (such as semantic graph
analysis), composed of many coordinating tasks operating on a data set so big
that it has to be stored on many physical devices. In such cases, it may be more 
efficient to dynamically choose where code runs as the applications progresses.

\end{abstract}

\section{Introduction}\label{sec:intro}
The emergence of distributed heterogeneous systems is driven by the ever
increasing demands for performance, energy efficiency, and cost reduction. For
example, in the last decade, the HPC community has been driving the adoption of
GPU as an accelerator for large-scale distributed systems and applications.
Recently, hyperscale service providers have introduced two new types of
datacenter infrastructure accelerators: the data processing unit (DPU) and the
computational storage drive (CSD). In contrast to GPUs, which have been
well-adopted by applications, both DPUs and CSDs are relatively new and have
very limited adoption. DPUs and CSDs are usually programmable devices that are
realized using FPGAs and/or Arm cores. Despite being designed with user
programmability in mind, these devices are typically exposed as fixed-function
components that provide transparent acceleration for a few popular usages, e.g.,
embedded Open vSwitch, IPSEC, and compression. As the list of available
functionalities are determined by datacenter vendors, applications are not
exposed to the programmable elements of DPUs and CSDs, and therefore cannot take
advantage of the devices' processing power for application specific purposes.

Developers are also challenged by the rapidly increasing amount of data they
have to deal with in their applications. For some applications the type and
distribution of the workload is highly dependent on the data and therefore
changes dynamically. Since moving data is still orders of magnitude slower than
doing computation, ideally we would like to move compute to data to improve
locality, not the other way around. Additionally, with new features being added
and tested on a daily basis, it could further slow down the development cycle if
the application needs to be re-compiled and re-deployed for every feature
addition and/or bug fix.

In this work, we aim to overcome the programmability barriers of such devices
by introducing the \ifunc{} API, an UCX API designed to facilitate the movement
of application-defined compute and data. Injected functions (\ifuncs{}), taking
the form of messages that contain binary code and data, are sent to and invoked
by other remote processes via the \cc{} framework~\cite{grodowitz2021cc}.
\ifuncs{} are similar to active messages in that each message contains data and
an action to perform, but their main difference is that \ifuncs{} actually
contain the code to be executed, while active messages contain only a reference
to the function to be called. \ifuncs{} provide more versatility because the
available functions are no longer fixed and a target system can register new
functions during run-time, without having to recompile UCX or the application.
Our API breaks the commonly used SPMD model of computation to benefit dynamic,
irregular, and data-driven applications on a wider range of heterogeneous
devices. The main contribution of this paper is an API and
implementation of RDMA-based remote function injection and linking.

The rest of this paper is organized as follows: \autoref{sec:back} provides an
overview on \cc{}, the high performance remote linking and messaging framework
leveraging the \ifunc{} API. This section also discusses some related work on
dynamic computation migration. Next, \autoref{sec:api} presents the \ifunc{}
API, how it could be used, how it is implemented, how it compares to UCX Active
Messages (AM), and how it can be secured. In \autoref{sec:eval}, we describe how
we validated our prototype implementation and discuss the initial benchmark
results. Finally, \autoref{sec:concl} provides our plans to continue to improve
the \ifunc{} API and \cc{} framework.
\section{Background}\label{sec:back}
The \ifunc{} API is an evolution of the remote function invocation mechanism of
the \cc{} framework. In this section we discuss the \cc{} framework as the
background of the work presented in this paper, as well as several related
works.

\subsection{\cc{}}
The original \cc{} framework is presented in~\cite{grodowitz2021cc}, which
covers implementation details of the framework and how its performance can be
improved using existing hardware features.

\cc{} is an extension of UCX~\cite{shamis2015ucx}, providing packaging,
transfer and execution of C functions in a fast and lightweight manner. It
aims at an API and a toolchain to enable the migration of compute and data
between local and remote CPU, GPU, DPU and CSD processes using UCX communication
capabilities. The users write the functions to be injected using a macro-based interface, then use the \cc{} toolchain to compile them into dynamic libraries
that can be loaded by the application at runtime. On the source process, the
executable code of the to-be-injected function is loaded from the dynamic
library, and is packaged with function arguments and a variable-length payload
to form a message (referred to as \textit{jams} in the original publication).

Upon receipt of a message containing injected functions, the target system
directly executes the C function embedded in the message. This mechanism employs
dynamic linking to support calling functions from libraries resident in the
target system from the injected functions.

\cc{} uses one-sided UCX put operations to enable fast delivery and execution of
injected functions. The runtime sets up a receiver thread waiting to call the
embedded function with minimal latency when a message arrives. For code in the
message to execute correctly on the receiver, the \cc{} toolchain statically
modifies the assembly to allow runtime linking against symbols on an arbitrary
host by redirecting all global offset table (GOT) accesses to an indirection
stored in the message. Remote runtime linking allows distributed application
updates to sub-processes of the application that alter their execution behavior
(without re-starting the process) by loading a library into a process to change
the resolution of objects or functions with fixed symbolic names. This way,
applications can implement dynamic control and compute with library loading and
linking.

\subsection{Related Work}
There are numerous libraries, frameworks, and runtimes that implement active
message semantics that have influenced the development of \cc{} and the \ifunc{}
API presented in this work. In brief, these projects include GASNet, Snap
Microkernel, Charm++, CHAMELEON, and FaRM.

GASNet~\cite{bonachea2018gasnet} is a communication library widely used on
high-performance computing clusters to implement advanced programming models.
In addition to normal data transfer routines, GASNet also provides a series of
APIs for registering and invoking active messages. GASNet uses the classical
function registration mechanism for identifying active message handlers, while
\ifuncs{} sends the executable code and does not require actions from the target
side. The Snap Microkernel~\cite{snap2019} project provides a platform for
remote procedure calls in the context of network functionality distribution.
Like many of the other computation placement and migration frameworks, it is a
heavyweight multifunction entity. Our \ifunc{} API has a smaller scope and could
be used as a building block as part of such a system. In the datacenter setting,
lightweight container launch for Lambda functions is implemented with
Firecracker~\cite{agache2020firecracker}. Another work from Fouladi et al.
provides very fast container launch to create highly granular lambda function
execution~\cite{fouladi2019laptop}. None of these projects addresses issues like
heterogeneity of hardware, since containerization is meant to abstract this.
\cc{} can be used as a shim between hardware and higher level libraries.

Charm++~\cite{acun2014parallel} implements distributed C++ objects with a
unified logical view of them (not partitioned to processes/ranks from the
programmer's perspective), plus the ability to call methods on those objects
regardless of their physical location, unlike regular active message where the
developer must decide when and where to request function invocations. The
Charm++ scheduler works behind the scene to distribute and migrate the objects
automatically, based on load distribution and communication patterns.
Its programming model runs at a very high level compared to \cc{} and UCX and
its runtime system supports lots of advanced features like fault tolerance. The
CHAMELEON~\cite{Klinkenberg2020Chameleon} framework by Klinkenberg et al.
uses compiler directives and runtime APIs to encapsulate OpenMP tasks as
migratable entities in a reactive workload balancer for irregular applications
written in MPI. Unlike CHAMELEON, \cc{} does not contain a load balancer, does
not depend on OpenMP or MPI or C++, nor requires explicit task progress if the
UCX library uses progress threads. Further, CHAMELEON's remote virtual address resolution
process to move tasks between address spaces is a heavyweight exchange of
references via MPI Send/Recv for each migration event. Our work could
potentially be used as a communication layer to greatly simplify and speed up
CHAMELEON, especially since they found in the course of their work that
push-oriented compute movement (as we have implemented here) is a better
mechanism than work stealing for load balancing since it allows
computation-communication overlap.

The FaRM~\cite{dragojevic2014farm} project implements a shared address space
programming model that uses the RDMA network for remote object manipulation.
\cc{} uses RDMA not only for moving data, it also injects user-defined functions
to remote machines using RDMA to provide higher flexibility while avoiding
re-compiling the application for functionality changes.

The \cc{} API developed in this paper builds on the semantics of the active
message API ~\cite{von1992active}, which combines a data payload with executable
code on a receiver. The primary innovation of the \cc{} API relative to
classical active message semantics is the ability to send binary function and
data payload simultaneously, without requiring the function to be present at
runtime compile time.
\section{Design and Implementation}\label{sec:api}
In this section, we present the design and implementation of the
\ifunc{} API, provide an example on the expected usage, and talk about its
security implications.

\subsection{The \ifunc{} API}

\begin{lstlisting}[caption=UCP \ifunc{} API, label={lst:ucp_ifunc_api},
                   style=my_style]
ucs_status_t
ucp_register_ifunc(ucp_context_h context,
                   const char*   ifunc_name,
                   ucp_ifunc_h*  ifunc_p)

void
ucp_deregister_ifunc(ucp_context_h context,
                     ucp_ifunc_h   ifunc_h)

ucs_status_t
ucp_ifunc_msg_create(ucp_ifunc_h      ifunc_h,
                     void*            source_args,
                     size_t           source_args_size,
                     ucp_ifunc_msg_t* msg_p)

void
ucp_ifunc_msg_free(ucp_ifunc_msg_t msg)

ucs_status_t
ucp_ifunc_msg_send_nbix(ucp_ep_h        ep,
                        ucp_ifunc_msg_t msg,
                        uint64_t        remote_addr,
                        ucp_rkey_h      rkey)

ucs_status_t
ucp_poll_ifunc(ucp_context_h context,
               void*         buffer,
               size_t        buffer_size,
               void*         target_args)
\end{lstlisting}

To start, the source process calls the \texttt{ucp\_register\_ifunc} function
with the \ifunc{}'s name \textit{ifunc\_name} to register an \ifunc{} library.
The UCX runtime will search the directory defined by the
\texttt{UCX\_IFUNC\_LIB\_DIR} environment variable for the dynamic library named 
\texttt{\textit{ifunc\_name}.so}, and uses \texttt{dlopen} and \texttt{dlsym} to
load the library and the user-provided \ifunc{} library functions defined in
\autoref{lst:ifunc_lib_api}, and finally returns a handler to the registered
\ifunc{}. Now \ifunc{} messages can be constructed using the
\texttt{ucp\_ifunc\_msg\_create} routine, which accepts user arguments and
passes them to the \ifunc{} library routines to prepare the \ifunc{}'s payload
that will be sent to the target process. Once the \ifunc{} message object is
created, it is ready to be written into the target process's memory using the
\texttt{ucp\_ifunc\_send\_nbix} routine, which uses the \texttt{ucp\_put\_nbi}
routine to write a continuous buffer into a memory region mapped by
\texttt{ucp\_mem\_map}.

On the target process, the \texttt{ucp\_poll\_ifunc} routine should be used to
wait on a UCP mapped memory region for incoming \ifunc{} messages. This routine
returns immediately if it could not find a newly received \ifunc{} message in
\texttt{buffer}. If a valid \ifunc{} message is found, the UCX runtime will
invoke the code contained in the \ifunc{} message with a pointer to the payload,
the size of the payload, and the \texttt{target\_args} pointer that points to
user-provided arguments on the target process. Currently, in our implementation,
the target process does not yet construct a GOT that contains redirections for
all the functions used by the \ifunc{} code, instead it uses the \ifunc{}'s name
contained in the message header to auto-register the specific \ifunc{} dynamic
library and uses the local GOT to patch the code shipped within the \ifunc{}
message. We plan to add GOT reconstruction functionalities in the future and the
target process will not need to register the \ifunc{} library anymore.

\begin{lstlisting}[caption=\ifunc{} library API, label={lst:ifunc_lib_api},
                   style=my_style]
void
[ifunc_name]_main(void*  payload,
                  size_t payload_size,
                  void*  target_args)

size_t
[ifunc_name]_payload_get_max_size(void*  source_args,
                                  size_t source_args_size)

int
[ifunc_name]_payload_init(void*  payload,
                          size_t payload_size,
                          void*  source_args,
                          size_t source_args_size)
\end{lstlisting}

A valid \ifunc{} library should define all three routines specified in
\autoref{lst:ifunc_lib_api}. The \texttt{[ifunc\_name]\_main} launches the
execution of the \ifunc{} code; it gets invoked when a \ifunc{} message is
received by \texttt{ucp\_poll\_ifunc} on the target process, with the three
arguments described in the previous subsection.

The \texttt{[ifunc\_name]\_payload\_get\_max\_size} and
\texttt{[ifunc\_name]\_payload\_init} routines are both invoked by the
\texttt{ucp\_ifunc\_msg\_create} routine on the source process. The first
routine is used by the UCX runtime to calculate the maximum size of the payload
to be sent within a \ifunc{} message for a given set of source process arguments
\texttt{source\_args}. Then the UCX runtime will allocate a \ifunc{} message
frame with a payload buffer of the requested size, and pass the same source
process arguments to the \texttt{[ifunc\_name]\_payload\_init} routine to
populate the payload buffer. This way, we eliminate unnecessary memory copies
while maintaining the flexibility of the interface.

\subsection{Using the API}

\begin{lstlisting}[caption=Sample \ifunc{} library, label={lst:sample_ifunc_lib},
                   style=my_style]
#include <paq8px.h>

size_t paq8px_payload_get_max_size(void *source_args,
                                   size_t source_args_size) {
    return est_output_size(source_args, source_args_size);
}

int paq8px_payload_init(void *payload,
                        size_t payload_size,
                        void *source_args,
                        size_t source_args_size) {
    encode(payload, payload_size,
           source_args, source_args_size);
    return 0;
}

void paq8px_main(void *payload,
                 size_t payload_size,
                 void *target_args) {
    db_handler dbh = target_args;
    decode_insert(dbh, payload, payload_size);
}
\end{lstlisting}

Here we provide an example on the expected usage of the \ifunc{} API. Suppose
the target process manages a database that stores voice recordings. When another
process wants to send a record compressed by the \texttt{paq8px} algorithm,
which is unsupported by the database, it can use the \ifunc{} library shown in
\autoref{lst:sample_ifunc_lib} to perform the task. The header file included at
the top of the library code contains the implementation of the algorithm, which
will be visible to the compiler during compilation. The first two user-provided
functions are used to encode and package payload on the source process, while
the main function performs payload decoding and database insertion on the target
process.

\begin{lstlisting}[caption=Sample \ifunc{} API usage, label={lst:sample_ifunc_usage},
                   style=my_style]
/* On the source process */
ucp_ifunc_h ih;
ucp_ifunc_msg_t msg;

ucp_register_ifunc(ucp_ctx, "paq8pv", &ih);
ucp_ifunc_msg_create(ih, record, record_size, &msg);

ucp_ifunc_send_nbix(ep, msg, recv_buffer, rmt_rkey);
ucp_ep_flush_nb(ep); // And wait on completion

ucp_ifunc_msg_free(msg);
ucp_deregister_ifunc(ucp_ctx, ih);

/* On the target process */
ucs_status_t s;
do {
    s = ucp_poll_ifunc(ucp_ctx, recv_buffer,
                       recv_buffer_size, db_handle);
} while (s != UCS_OK);
\end{lstlisting}

During run-time, as demonstrated by \autoref{lst:sample_ifunc_usage}, the source
process registers the \texttt{paq8px} library, constructs an \ifunc{} message
with the recording as its payload, and sends it to the target process. On the
target process, the polling loop calls the \texttt{ucp\_poll\_ifunc} function
until it returns \texttt{UCX\_OK}, which indicates that it has received and
executed an \ifunc{} message. If the user would like the target process to poll
for incoming \ifunc{} messages continuously, the \texttt{ucs\_arch\_wait\_mem}
routine can be used to wait on memory locations that \ifunc{} messages are
expected to arrive and reduce resource usage.

\subsection{\ifuncs{} \textit{versus} UCX Active Messages}
Injected functions are inspired by active messages but are different in many
aspects. A comparison between \cc{} injected functions and UCX active messages
helps the reader know the differences and decide which one to use.

We start by listing the main similarities. \ifuncs{} and UCX AMs allow sending
payloads of various sizes and launching functions on remote processes. Both
accept user-defined arguments when the functions are launched on the target
processes, so the functions have access to resources in the local address space.
Lastly, both mechanisms require active progression on the target side to process
the received messages, in the form of non-blocking polling calls.

The main difference between active messages and injected functions is that,
instead of establishing a mapping between registered functions and unique IDs,
\ifunc{} messages carry the actual binary code of the functions and the functions
themselves are identified by a name. This key \ifunc{} feature enables a set of \ifunc{}
benefits over UCX AMs. The first of these benefits is that \ifuncs{} can be
loaded on-demand during run-time, without recompiling the application; AM
handlers are determined at compile time, requiring the application to be stopped
and recompiled when AM handlers are added or modified. A related benefit to this
one is that, since the function code is sent with each invocation, the code can
be modified anytime under the same \ifunc{} name. Another \cc{} difference is
that \ifuncs{} are registered on the source process while AM handlers are
registered at the target process; this feature of the \ifunc{} API enables the
system to dynamically add nodes with no previous knowledge of what functions it
might need to execute in the future.

UCX AMs use
on-demand internal buffers for receiving messages, while \ifuncs{} require the
user to allocate special buffers and a consensus about where the target
processes expect the messages to arrive. \ifuncs{} need special modifications to
the assembly code before they can be used, while AM code does not. We expect
these limitations to go away as we keep improving the \cc{} framework.

\subsection{Implementing the API}

\begin{figure}
    \includegraphics[width=\textwidth]{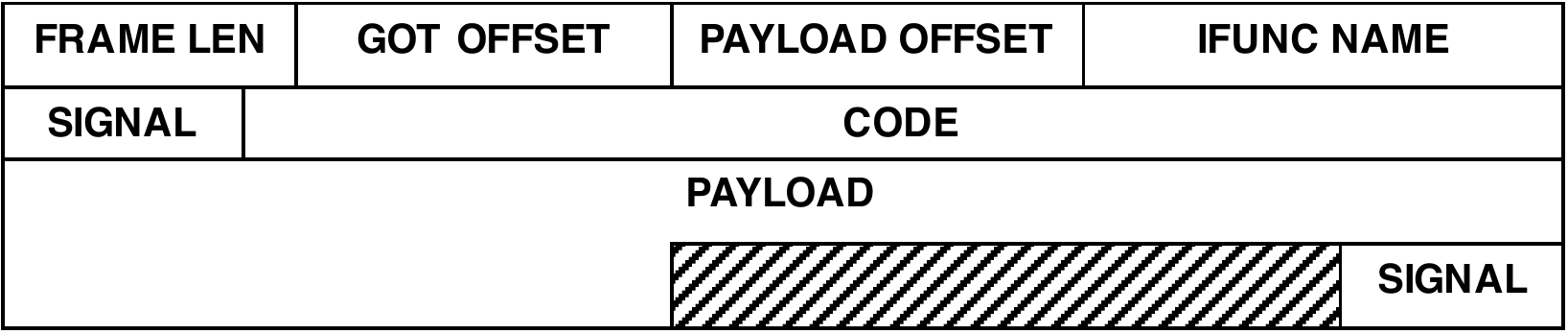}
    \caption{Structure of an \ifunc{} message} \label{fig:frame}
\end{figure}

Each \ifunc{} message, constructed by the \texttt{ucp\_ifunc\_message\_create}
routine, is composed of a header, a code section, an optional payload section,
and a trailer signal, as seen on \autoref{fig:frame}. If the code section is a
direct copy of the \texttt{.text} section of the \ifunc{} dynamic library,
external function calls (e.g. \texttt{printf}) and accesses to global variables
will not have the correct relocations on the target process, due to Linux's
relative addressing and address space layout randomization (ASLR). To fix this
issue, we compile the \ifunc{} dynamic library with the \texttt{-fno-plt} flag
to force all relocations to go through the global offset table (GOT), skipping
the procedure linkage table (PLT). Then we use a Python script to modify the
assembly code so that all references to the GOT will redirect through another
table on the target process. A pointer to this alternative table is inserted as
a hidden global variable by the script and is shipped with the \ifunc{} message
as part of the code, and the target process is expected to fill this variable
with the address of a reconstructed GOT before invoking the \ifunc{}'s main
function.

\begin{figure}
    \includegraphics[width=\textwidth]{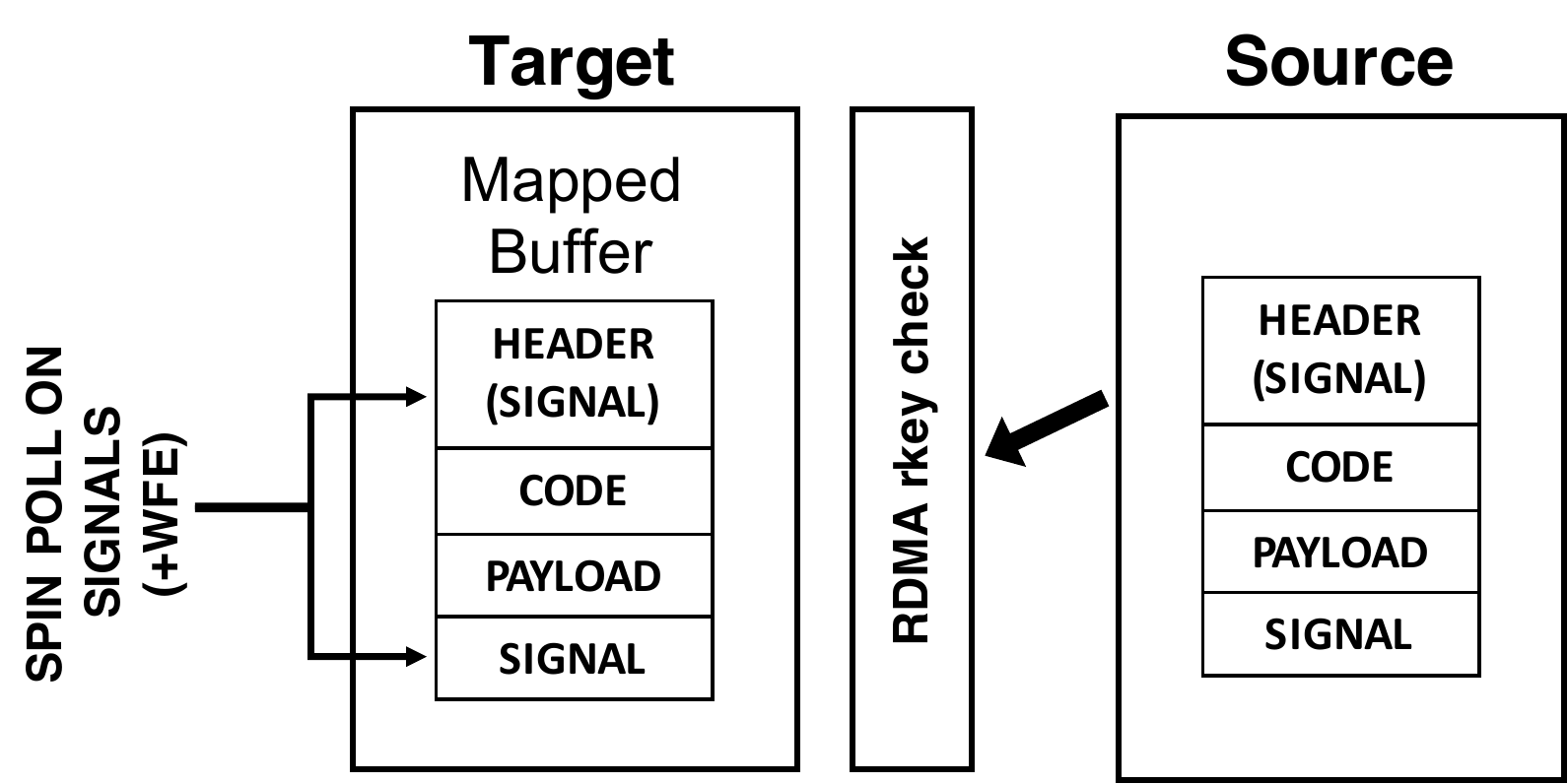}
    \caption{\ifunc{} source-target communication} \label{fig:comms}
\end{figure}

When an \ifunc{} message arrives at the target process, the integrity of the
header is verified using the header signal, and messages that are ill-formed or
too long will be rejected. Then the runtime parses the header to get the total
size of the message frame and waits for the trailer signal to arrive, as shown
in \autoref{fig:comms}. In our tests, we use the \texttt{WFE} instruction to
reduce resource usage when busy-waiting on the trailer signal, without incurring
a heavy performance penalty.

Before calling the main function of a fully delivered \ifunc{} message, the
target process should perform work similar to a dynamic linker: construct a GOT
that has all the relocations needed by the \ifunc{} code in the correct offsets.
This mechanism is not implemented yet. Instead, we assume the same \ifunc{}
dynamic library is also available on the target process's file system, so the
target process can simply load the library and let the system dynamic loader do
the GOT construction. In our implementation, the \texttt{ucp\_poll\_ifunc}
routine uses the \ifunc{}'s name provided by the message header to attempt the
auto-registration of any first-seen \ifunc{} type. If the corresponding library
is found and loaded successfully, the UCX runtime will patch the alternative GOT
pointer of the code section of the \ifunc{} message with a pointer to the same
library's GOT in the local address space, and store the related information in a
hash table for subsequent messages of the same type. We plan to implement the
dynamic linking and GOT reconstruction mechanism in the future.

\subsection{Security Implications and Mitigations}\label{sec:security}
A full security model design and implementation is well beyond the scope of this
paper. This section provides an overview of security challenges and directions
for security improvements.

For our \cc{} framework implementation, we have relied on the built-in security
mechanisms defined by the UCX framework and the IBTA
standard~\cite{ibta-specification}, which underpins RDMA interconnects.
Specifically, we are using a remote access key (RKEY) to register and control
remote memory accesses. For IBTA interconnects, the RKEY is defined as a
32-bit value. When the memory is registered for remote memory access, the
underlying interconnect generates the RKEY based on a virtual memory address and
the permissions (remote read, write, or atomic access). In order to access the
memory region over the RDMA interconnect, the target process has to provide the
RKEY to the RDMA initiator through an out-of-band channel. Then, the remote
memory access initiator uses the RKEY to remotely read and write to the target
process memory. If the process accesses the memory with an invalid RKEY, the
request gets rejected at the hardware level.

There are a number of security concerns~\cite{taranov-redmark} regarding the
strength of RKEY protection as defined by the IBTA standard. Improvements to the
IBTA security model are out of scope for this work. However, since we have
constructed this as a module of the UCX framework, the implementation is not as
strictly tied to the IBTA network implementation.
\section{Evaluation}\label{sec:eval}
In this section, we present our test and evaluation efforts, along with testbed
and benchmark descriptions. We end the section by showing and analyzing the
initial results.

\subsection{Microbenchmark Description}
To verify and do a preliminary evaluation of our API and its implementation, we
ran message throughput and ping-pong latency benchmarks with a simple \ifunc{}
library and we compare them against the same benchmarks written using UCX AM.
The results are presented below. In both benchmarks, the \ifunc{} main function
simply increases a counter on the target process used to count the number of
executed messages.

In the \ifunc{} message throughput benchmark, a ring buffer is allocated using
the \texttt{ucp\_mem\_map} routine so it allows UCP put operations. The source process
fills the buffer with \ifunc{} messages of a certain size, flushes the UCP
endpoint used to send the messages, then waits on the target process's
notification indicating that it has finished consuming all the messages before
continuing to send the next round of messages. This leads to some overhead but
is not significant when the number of messages is large. For the equivalent UCX
AM throughput benchmark, since the UCX runtime uses internal buffers to handle
the messages, the source process simply sends all the messages in a loop and
flushes the endpoint at the end.

The ping-pong benchmark is implemented using the classical approach: each
process sends a message, flushes the endpoint and waits for the other process to
reply before continuing this process.

\subsection{Testbed Platform}
The development and evaluation testbed for this work consisted of two servers,
each with a 4-core, Arm-based modern superscalar processor with a 1MB dedicated
L2 cache per core, a 1MB shared L3 cache per 2-core cluster, and a 8MB shared
last level cache (LLC). The core clock is 2.6GHz and the on-chip interconnect
clock is 1.6GHz. Each server has 16GB of DDR4-2666 main memory. For the
interconnect, we used
two Mellanox/Nvidia ConnectX-6 200Gb/s InfiniBand dual-port HCAs. The two
systems were connected back-to-back (no InfiniBand switch) using the first port
on each ConnectX-6 HCA. The servers used Ubuntu 20.04, running a custom
Linux 5.4 kernel, modified to allow user space control of the CPU prefetching
mechanisms. We used the RDMA and InfiniBand drivers that came with Mellanox
OFED, versioned \texttt{OFED-5.3-1.0.0d}.

\subsection{Experimental Results and Analysis}
\begin{figure}
    \includegraphics[width=\textwidth]{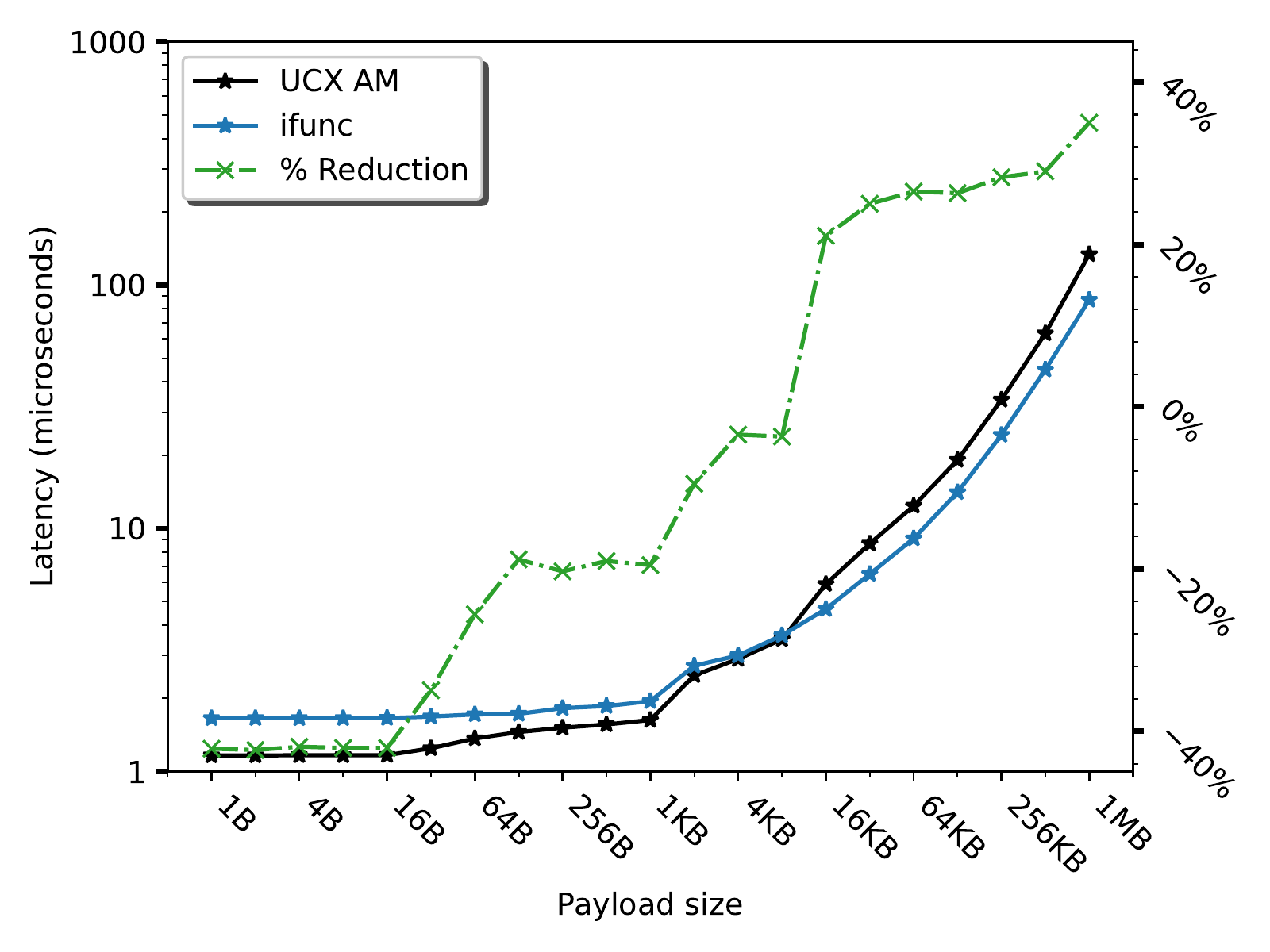}
    \caption{Latency comparison between \ifunc{} and UCX AM, including \ifunc{}
    latency reduction with respect to UCX AM latency} \label{fig:lat}
\end{figure}

\autoref{fig:lat} shows the one-way latencies of sending and executing the
benchmark function using the \ifunc{} and UCX AM APIs. For smaller payloads, the
\ifunc{} latency is up to 42\% slower than the AM latency. As payload (and
message) size increases, the \ifunc{} latency gets closer to that of AM,
crossing over somewhere between payload sizes 8KB and 16KB. After this crossover
point, the \ifunc{} latency keeps improving, reaching a 35\% latency reduction
for the 1MB payload size. For small payload sizes, we expected the AM latencies
to be better because the code sent in the \ifunc{} messages dominate the message
size, not the payload. That being said, the performance gap is larger than it
needs to be because of the \texttt{clear\_cache} operation.

To ensure the correct operation of the \ifunc{} invocation, the instruction
cache needs to be cleared after the runtime confirms the data has arrived
because the I-cache could have stale data due to some systems
not having coherent I-caches. \texttt{glibc}'s Arm64 \texttt{clear\_cache}
implementation avoids clearing the I-cache when it detects a coherent I-cache by
reading an architectural register. Our testbed did not have a
coherent I-cache, and that is why the arrival of each \ifunc{} incurs a
performance hit on the target system. This is likely to be the reason why the
latencies were not better.

\begin{figure}
    \includegraphics[width=\textwidth]{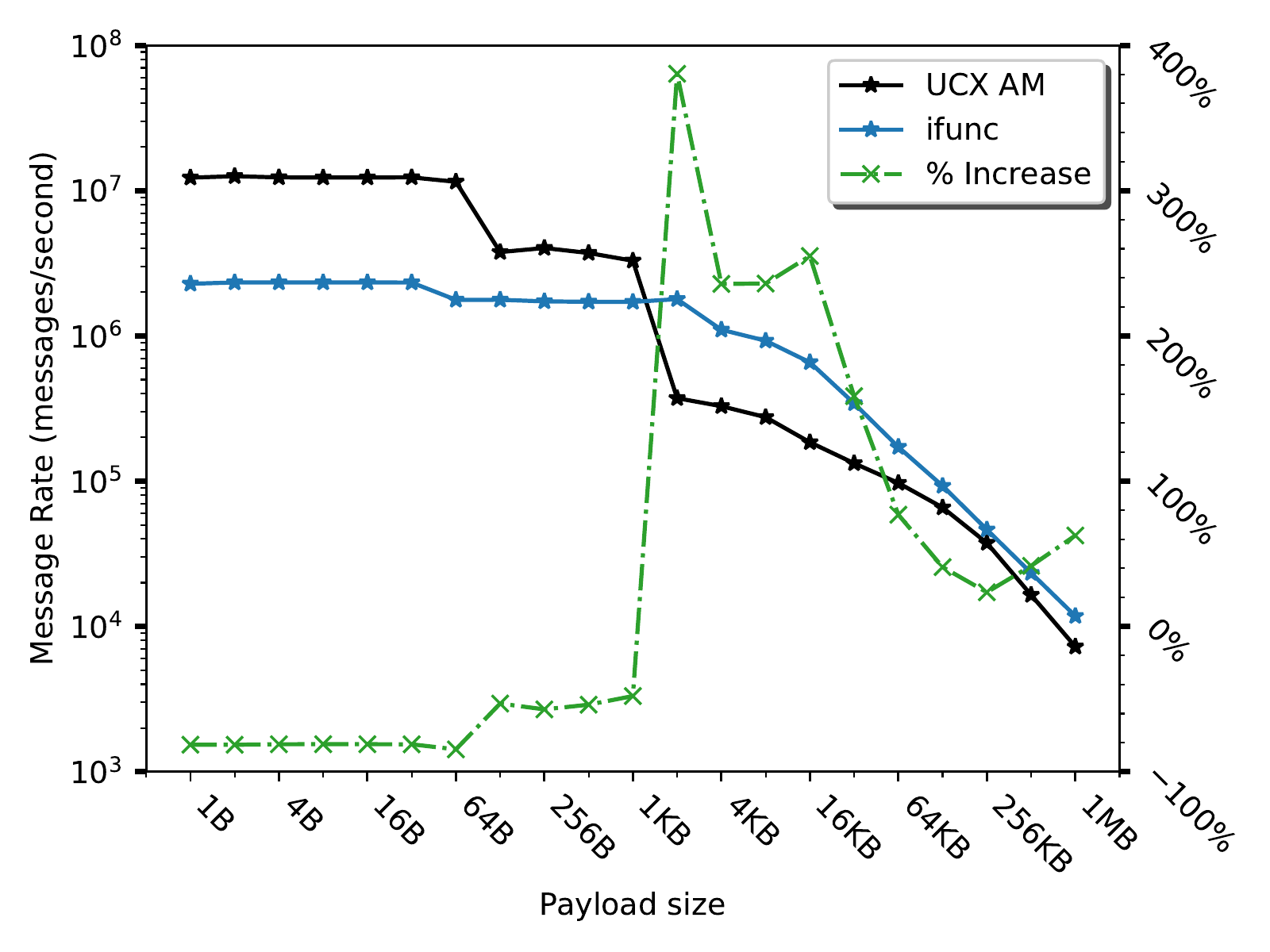}
    \caption{Message throughput comparison between \ifunc{} and UCX AM,
    including \ifunc{} throughput increase with respect to UCX AM message
    throughput} \label{fig:mr}
\end{figure}

\autoref{fig:mr} presents the results of the message throughput benchmarks using
both APIs. For 1B payloads, the \ifunc{} message rate is 81\% lower than that of
UCX AMs. The message rate continues to be worse until the payload goes from 1KB
to 2KB. From this point on, the \ifunc{} message rate is superior, first spiking
at 380\% better, then dropping to 23\% and then coming back up to 62\% higher.

One interesting observation is the \textit{stepping} experienced by the UCX AM
line. These steps are likely due to the change is underlying protocol for moving
the active messages. Interestingly, the point where \ifuncs{} start performing
better coincides with the sharp performance falloff \textit{step} experienced by
UCX AM, possibly due to protocol differences between \ifunc{} and UCX AM.

As in the latency case, we think that the \ifunc{} performance would have been 
better if we had evaluated using a platform with a coherent I-cache. Another
area where we could have extracted more performance is the buffer mechanism used
to send messages to the target.

\subsection{Takeaways}
From these initial benchmarks, we observe that the \ifuncs{} perform worse than
UCX Active Messages for small payload sizes. The larger the payloads become, the
better \ifuncs{} behave. This small-payload behavior is expected because, while
active messages carry a numerical ID alongside the payload, \ifuncs{} actually
carry the function binary alongside the payload. Despite the \ifunc{} messages
being larger, we expected them to be more performant, but the slowdowns could be
explained by the need to perform a \texttt{clear\_cache} operation on the
instruction cache because our testbed does not have a coherent I-cache.

Since this is a preliminary evaluation of the \ifunc{} API, we plan need
to run additional benchmarks to better understand the behavior of the \cc{} framework
with a wider set of micro-benchmarks and applications in the future.
\section{Conclusion}\label{sec:concl}
The \ifunc{} API and \cc{} framework provide a high performance mechanism
of moving compute and data over networks between a wide class of processing
elements. It uses dynamic linking and loading to resolve \ifunc{} external
symbolic references on a per-process basis. We presented the user-facing side of
the API and how developers write \ifunc{} libraries. Important differences
between \ifuncs{} and traditional active messages were also discussed. We
provided an overview of the security mechanisms of that \cc{} can leverage. The
code is released on GitHub~\cite{two-chains-source}. We will discuss the future
steps on the evolution of \cc{} and the \ifunc{} API below.

\subsection{Future Work}\label{sec:future}
Our \cc{} vision does not require the presence of the \ifunc{} dynamic library
on the target's filesystem. We implemented it first this way in our prototype
because it was going to allow us to have a version working sooner for
evaluation. We are looking into ways of removing this requirement so \cc{}
target processes are able to handle received \ifunc{} messages with the correct
dynamic linking mechanism.

We are also working on switching the underlying implementation of \cc{} to use
UCX's send-receive semantics instead of RDMA Puts. This change will enable a
simpler API because the user would not have to worry about setting up a
RWX-enabled buffer on the target process. In addition, the user would not have
to tell the source process exactly where to PUT the messages. This change would
also eliminate the need for a special polling API and calling it from the target
process to process incoming messages, as \ifuncs{} will be progressed with
other UCX operations by calling \texttt{ucp\_worker\_progress}. The good thing
with this change is that the current API will only have minimal changes: we
would mostly be removing unnecessary arguments and functions calls.

Currently the payload is tightly packed after the code segment of the \ifunc{}
message frame so we do not have any data alignment guarantees. This could be
undesirable for vectorization and some other applications. We plan to allow the
user to specify an alignment requirement on the payload buffer to better support
vectorization and other needs.

The compilation toolchain of this work uses a Python script to prepare the
\ifunc{} code section to accept a patched GOT. We are considering updating the
way we do this to make this important step target-process-architecture agnostic.

We are still debugging and stress-testing the \cc{} and its API implementation.
We are also working on getting \cc{} in a state where it can be accepted to
upstream UCX. Finally, we will test the \cc{} framework with benchmarks that
do useful work and on a machine that has a coherent I-cache.

\subsubsection{Acknowledgments.}
The authors would like to thank the Los Alamos National Laboratory for their
continued support of this project. In addition, we would like thank Curtis
Dunham, Megan Grodowitz, Jon Hermes, and Eric Van Hensbergen for their review of
the paper and code.

%

\bibliographystyle{splncs04}
\bibliography{\jobname}

\end{document}